\documentclass[aps, reprint, floatfix, amsmath, amssymb]{revtex4-2}
\usepackage{float} 
\usepackage{graphicx}
\usepackage{dcolumn}
\usepackage{bm}
\usepackage{xcolor}
\usepackage{xfrac}
\begin{document}

\preprint{APS/123-QED}

\title{SPEA - an analytical thermodynamic model for defect phase diagram}

\author{Jing Yang}
\email{j.yang@mpie.de}
\affiliation{Computational Materials Design Department, Max Planck Institute for Sustainable Materials, Max-Planck-Str. 1, D-40237 Düsseldorf, Germany}

\author{Ahmed Abdelkawy}
\affiliation{Computational Materials Design Department, Max Planck Institute for Sustainable Materials, Max-Planck-Str. 1, D-40237 Düsseldorf, Germany}

\author{Mira Todorova}
\email{m.todorova@mpie.de}
\affiliation{Computational Materials Design Department, Max Planck Institute for Sustainable Materials, Max-Planck-Str. 1, D-40237 Düsseldorf, Germany}
 
\author{Jörg Neugebauer}%
\affiliation{Computational Materials Design Department, Max Planck Institute for Sustainable Materials, Max-Planck-Str. 1, D-40237 Düsseldorf, Germany}

\date{\today}

\begin{abstract}
We propose an analytical thermodynamic model for describing defect phase transformations, which we term the statistical phase evaluation approach (SPEA). The SPEA model assumes a Boltzmann distribution of finite size phase fractions and calculates their statistical average. To benchmark the performance of the model, we apply it to construct binary surface phase diagrams of metal alloys. Two alloy systems are considered: a Mg surface with Ca substitutions and a Ni surface with Nb substitutions. To construct a firm basis against which the performance of the analytical model can be leveled, we first perform Monte Carlo (MC) simulations coupled with cluster expansion of density functional theory dataset. We then demonstrate the SPEA model to reproduce the MC results accurately. Specifically, it correctly predicts the surface order-disorder transitions as well as the coexistence of the 1/3 ordered phase and the disordered phase. Finally, we compare the SPEA method to the sublattice model commonly used in the CALPHAD approach to describe ordered and random solution phases and their transitions. The proposed SPEA model provides a highly efficient approach for modeling defect phase transformations.
\end{abstract}

\maketitle

\section{Introduction}

Thermodynamic phase diagrams are an important tool for materials engineering as they provide the basis for understanding the composition-structure-property relationship of materials. A powerful and widely used approach to construct phase diagrams is the CALculation of PHAse Diagrams (CALPHAD) method. The CALPHAD method is based on carefully collected thermodynamic data of bulk systems and provides analytical models with empirical fitting parameters~\cite{Kaufman1970, Hillert2007}. Today, the thermodynamic databases and the CALPHAD models to construct phase diagrams are commercially available~\cite{Andersson2002,Bale2009} and are widely used in materials design~\cite{Kaufman2014, Luo2015, Miracle2017, Agren2023}. 

Parallel and largely independent to the progress of CALPHAD for bulk phases, the concept of phase diagrams has also been developed by the surface science community. The surface phase diagrams are constructed using {\it ab initio} calculations and allow for identifying the stable surface composition, structure, and solute/adsorbate configuration at given thermodynamic condition~\cite{Northrup1993, Felice1996, Reuter2001, VandeWalle2002}. More generally, such phase diagrams can be used to describe the reconstruction and segregation behavior at 1D or 2D structural defects, including surfaces, interfaces, grain boundaries, and dislocations. These phase diagrams are referred to as defect phase diagrams~\cite{VandeWalle2002, Turlo2019, Kerzel2022, Starikov2023, Tehranchi2024, Matson2024}. They provide direct insights into the relationship between materials properties and microstructure, and open up new opportunities and perspectives in materials engineering.


With the development of \textit{ab initio} calculations, \textit{ab initio} thermodynamics has been gaining importance in the construction of (bulk) phase diagrams \cite{Liu2009, Liu2020, Chew2023, Liu2023}. They serve as useful benchmarks against experimental uncertainties and provide additional information on the unstable and metastable phases. Accurate and systematic datasets have become available with the advent of finite temperature \textit{ab initio} calculations and high throughput simulations \cite{Zhu2017, Grabowski2019, Zhu2020, Menon2021, Zhu2024}. More recently, computational thermodynamics has been further expanded by the development of \textit{ab initio} based machine learning interatomic potentials, enabling computationally-efficient large-scale modeling in the multi-component space~\cite{Zong2018, Drautz2019, Deringer2019, Mishin2021, Rosenbrock2021, Santos-Florez2023, Kotykhov2023}. Fast and automated computation of phase diagrams with density functional theory (DFT) accuracy is on the horizon~\cite{Menon2024}. The computation of defect phase diagrams, however, poses greater challenges as compared to their bulk counter parts due to the complex interplay between the microstructure and defect states. With the progress in computation power and efficiency, high-throughput computation of structural defects are becoming accessible. To describe these structurally and configurationally highly complex defect diagrams, thermodynamic models that go beyond the existing CALPHAD method are required.

\begin{figure}[t] 
\centering
\includegraphics[width=0.45\textwidth]{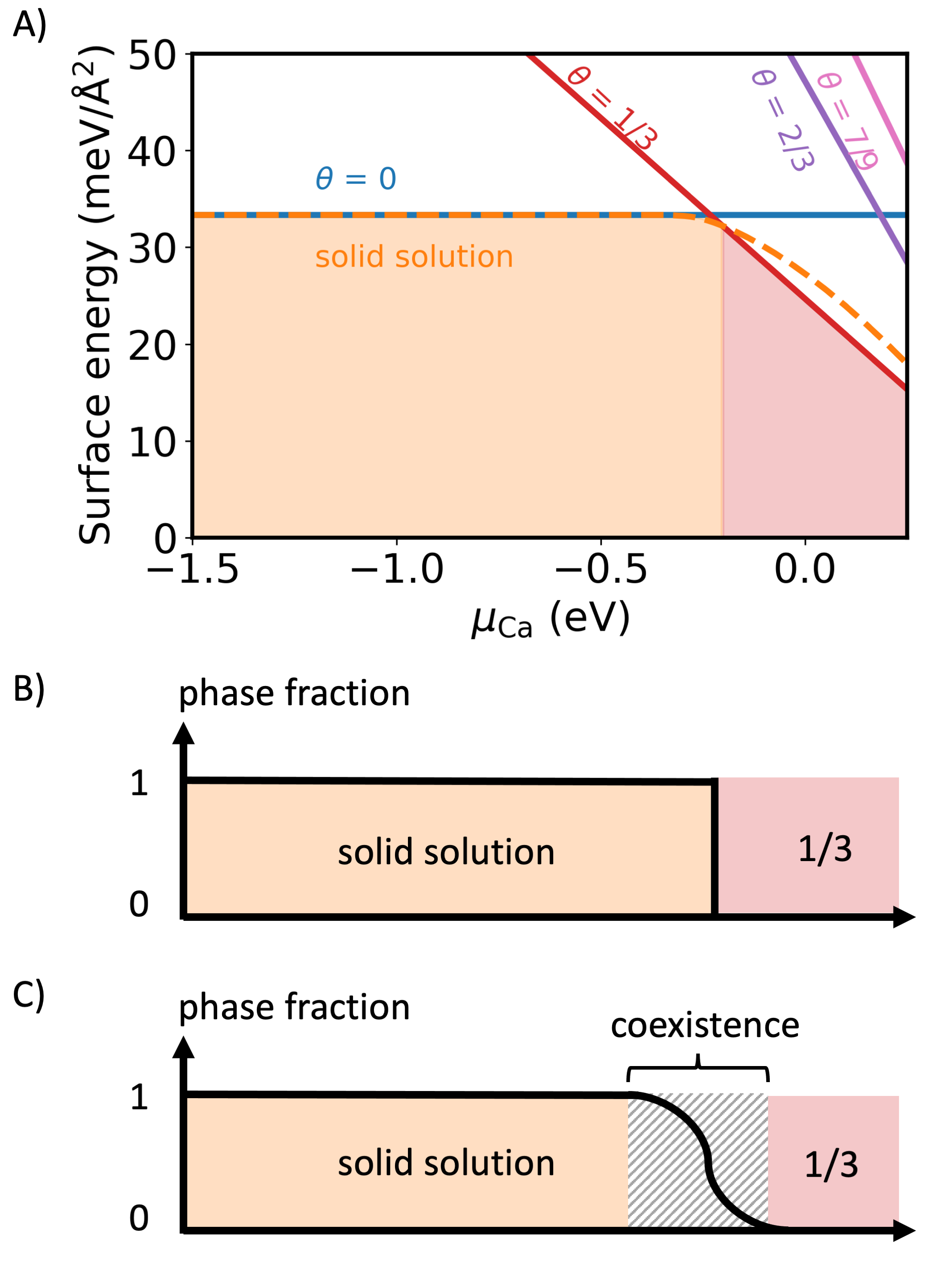}
\caption{\label{fig:occupation} (a) A surface phase diagram of Mg(0001) surface with Ca substitution in vacuum, adopted from Ref. \citenum{Yang2023}. (b) The schematic showing the phase fraction of the solid solution phase of the conventional interpretation, where the transition between phases is abrupt. (c) The schematic showing the realistic phase fraction change, where a coexistence region exists at the phase boundary.}
\end{figure}

In this work, we consider surface phase diagrams as prototype for generic defect phase diagrams. Specifically, we focus on describing the surface phase transformation on alloy surfaces. In conventional surface phase diagrams, the lowest energy phase is shown for the given thermodynamic condition, assuming a phase fraction of 1 (as examplified in Fig. ~\ref{fig:occupation}). This corresponds to the commonly employed concept that the surface phases cannot coexist and phase transformations thus happen abruptly. We perform Monte Carlo (MC) simulation based on a density functional theory (DFT)-cluster expansion (CE) model and show on two alloy systems, Mg alloyed with Ca and Ni alloyed with Nb, that phase mixture occurs in the transition zone. Based on this insight, we then propose a Statistical Phase fraction Evaluation Approach (SPEA), which calculates the area fraction of each defect phase according to Boltzmann distribution. To test the performance of this model against existing analytical models for bulk thermodynamics, we compare the proposed SPEA model with the sublattice model, a model which is commonly used in CALPHAD modeling.   


\section{Method}

\subsection{Density Functional Theory Calculation}

DFT calculations are performed in a plane wave based DFT framework and the projector augmented wave (PAW) approach as implemented in the Vienna Ab-initio Simulation Package (VASP)~\cite{Kresse1993, Kresse1996}. Exchange and correlation are described with the PBE implementation of the Generalized Gradient Approximation (GGA)~\cite{Perdew1996}. We use Fermi smearing of 0.1\,eV, converge energies to $10^{-5}$\,eV and optimize geometries using a force criterion of 0.01\,eV/{\AA}. An energy cut-off of 500 eV is used. For the Ni-Nb system, we perform spin-polarized calculation with a ferromagnetic ordering of the Ni metal. 
Surfaces are modelled using six layer thick slabs and a vacuum region of 12\,{\AA}. We consider ($2 \times 2$) and ($3 \times 3$) surface unit cells with all possible permutations of the solute configuration. For the ($2 \times 2$) cell, we use a $k$-point mesh of ($6 \times 6 \times 1$) and for the ($3 \times 3$) cell  ($4 \times 4 \times 1$). These parameters ensure the convergence of the surface energy to 10 meV per (1$\times$1) surface unit cell. For the Mg-Ca calculation, we use the pseudo-potential with 2 valence electrons. For the Ni-Nb calculation, we use the PAW pseudo-potentials with 16 valence electrons for Ni and 13 valence electrons for Nb. 

\subsection{Semi-grand Canonical Monte Carlo}

We run semi-grand canonical Monte Carlo (SGCMC) simulations based on cluster expansion~\cite{Sanchez1984}. The cluster expansion is obtained from fitting to DFT surface energy calculations \cite{Yang2023}. The surface energy is defined as
\begin{equation}
    E^{\rm surf}_{\sigma}(\{\mu_{i}\})  = \frac{1}{2}[E_{\sigma}^{\rm DFT} - \sum_{i = \rm A, X}n_{i,\sigma}  (E_{i\text{-}\rm bulk}^{\rm DFT} + \mu_{i})  
     ].
\label{eq:surf_energy_general_expression}
\end{equation}
Here $E_{\sigma}^{\rm DFT}$ is the DFT-calculated energy of a surface slab, where the subscript $\sigma$ represents different surface configurations. A and X represent the host metal and the solute species, respectively. $n_{i,\sigma}$ represents the number of atoms of species $i$ in the surface cell,  $E_{i\text{-}\rm bulk}^{\rm DFT}$ the DFT-calculated energy of the bulk metal of species $i$, and $\mu_i$ the chemical potential. The factor 1/2 is due to the use of a symmetric slab. With this definition, the chemical potential of the bulk metal is $\mu_i = 0$.

In the semi-grand canonical ensemble, the total number of atoms is fixed, but the surface is free to exchange atoms with an A-X reservoir. In the reservoir, the chemical potential of element A is $\mu_{\mathrm{A}}$, representing the bulk metal. The chemical potential of X is calculated from the concentration $x_{\rm X}$of X in the bulk of A, assuming the dilute limit:
\begin{equation}
      \mu_{\rm X\text{-}in\text{-} A\text{-}bulk}(x_{\rm X}, T) = \Delta E_{\rm X\text{-}in\text{-} A\text{-}bulk} + k_B T \ln x_{\rm X}. 
      \label{eq:chemical_potential}
\end{equation}
Here $\Delta E_{\rm X\text{-}in\text{-} A\text{-}bulk}$ is the DFT-calculated energy of having one X atom in the bulk of metal A
\begin{equation}
    \Delta E_{\rm X\text{-}in\text{-}A\text{-}bulk} = E^{N-1}_{\mathrm{A} + 1 \mathrm{X}}-\frac{N-1}{N}E^{N}_{\mathrm{A}}-E_{\rm X\text{-}bulk},
\end{equation} 
where $E^{N}_{\mathrm{A}}$ is the DFT energy of bulk metal A with $N$ atoms, $E_{\rm X\text{-}bulk}$ the DFT energy of one atom of bulk metal X, and $E^{N-1}_{\mathrm{A} + 1 \mathrm{X}}$ the DFT energy of one X atom substitution in the bulk metal A matrix with $N$ atoms. In the Mg-Ca system, $E_{\mathrm{Ca\text{-}in\text{-}Mg\text{-}bulk}}$ = 95 meV. In the Ni-Nb system, $E_{\mathrm{Nb\text{-}in\text{-}Ni\text{-}bulk}}$ = $-$0.646 eV. In the SGCMC simulation, only the difference between the two chemical potentials $\delta \mu = \mu_{\mathrm{A}}-\mu_{\mathrm{X}}$ affects the simulation result, not their absolute values. $\delta \mu$ represents the energy cost of exchanging one atom in the bulk of metal A from A to X. 

For each SGCMC simulation, we run 2$\times$10$^7$ MC steps and determine the equilibrium coverage by taking the average coverage of the last 10$^6$ steps of the run, which is long enough to ensure convergence. We also test the effect of surface cell size on the obtained surface coverage isotherm [see Supplemental Material (SM) \cite{SM}]. In the following, we report the result of using a (60$\times$60) cell. We observe negligible changes when going to a larger surface size.

\subsection{The Free Energy of the Disordered Phase}

\begin{figure}[t] 
\centering
\includegraphics[width=0.5\textwidth]{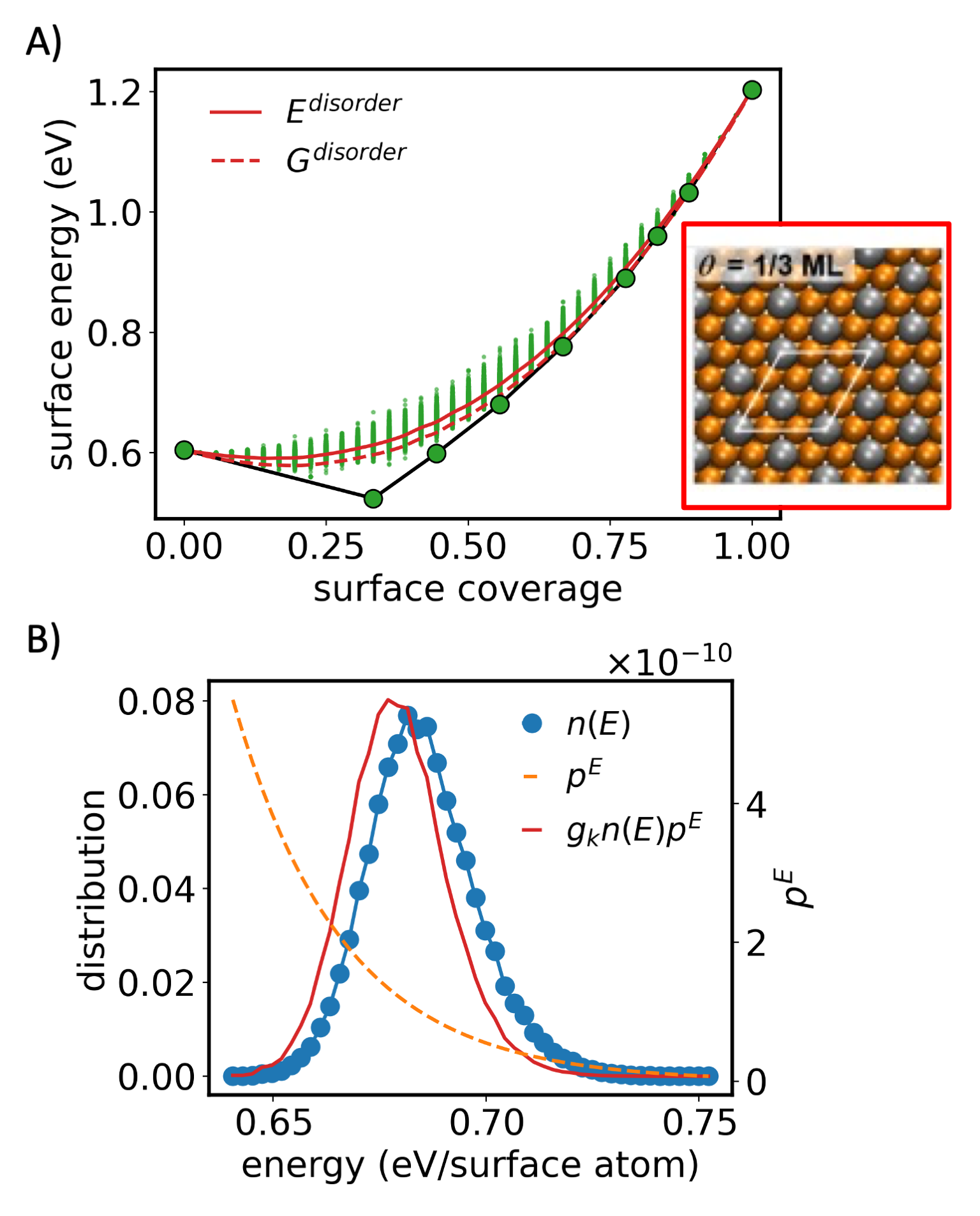}
\caption{\label{fig:disorder} (a) Energy distribution of the Mg-Ca surface with different solute configurations as a function of the surface solute coverage $\theta$ predicted by cluster expansion (CE). The surface energy is shown for the (1$\times$1) surface unit cell. Here the surface energies are calculated at the chemical potential of bulk Mg and Ca metal. The surface atomic structure on the right represents the surface configuration with the lowest energy ($\theta = 1/3$\,ML ordered phase), with the orange and silver spheres representing Mg and Ca, respectively. The $E^{disorder}$ and $G^{disorder}$ lines are the energy and free energy of the disordered phase at 300\,K calculated using the thermodynamic model described in Sec. IIC. (b) The energy distribution of different surface states at $\theta = 0.5$\,ML (blue line), the Boltzmann probability distribution $p^E$ of each state at the given energy $E$ (orange line), and the corresponding probability distribution at 300\,K (red line).   }
\end{figure}

Defect phase diagrams such as the one shown in Fig. \ref{fig:occupation} naturally describe the chemically ordered defect phases. Since our aim is to include also the order-disorder transition, our thermodynamic model has to include also the chemically disordered defect phase. It is therefore important to obtain the free energy of the disordered surface phase. However, the computational cost and the limitation of the simulation cell size make it impractical to directly model the disordered phase using DFT calculation. To this end, we use the cluster expansion approach to sample a large number of different permutations of the solute atom arrangement using a (6$\times$6) surface cell. Fig. \ref{fig:disorder}a shows the sampled configurations at different surface coverages (green dots). The black line shows the convex hull. Each point on the convex hull corresponds to an ordered phase. 

At each given surface coverage $\theta$, we consider the energy of the disordered phase $E_{\theta}$ and its entropy $S_{\theta}$. $S_{\theta}$ is given by 
\begin{equation}
    S_{\theta} = -k_B\sum_{i} p_i \ln p_i, 
\end{equation}
where the sum over $i$ denotes all possible permutations of surface solute atoms at concentration $\theta$ and $p_i$ is the probability of each state. $k_B$ is the Boltzmann constant. Given that the total number of atoms in the surface cell is $N$ and the number of solute atoms is $k$, the total number of possible states is $g_k = C_N^k$. The energy is
\begin{equation}
    E_{\theta} = \sum_{i} p_i E_i.
\end{equation}

One can first consider the case when all states have the same energy, i.e. $E_i$ is a constant, and the probability of each state $i$ is $p = 1/g_k$. For this case the model falls back to ideal mixing ($S_{\theta} = -k_BT[\theta\ln \theta + (1-\theta)\ln(1-\theta)]$) for sufficiently large $N$. In contrast to the ideal case, in Fig. \ref{fig:disorder}a we observe a spread in the distribution of the energies for a given surface coverage. As a consequence, thermodynamic quantities such as surface energies and entropies deviate from the ideal mixing model. In Fig. \ref{fig:disorder}b we show the density of states of the energy distribution $n(E)$ for $\theta = 0.5$\,ML.

In Fig. \ref{fig:disorder}, we denote the probability of a state with energy $E$ as $p^E$. From the normalization of the probability, we have
\begin{equation}
     \sum_E g_kn(E)p^E(T) = 1,
\end{equation}
i.e. the sum of all the individual state probabilities is 1. Assuming that $p^E$ follows the Boltzmann distribution, we have
\begin{equation}
    p^E(T) = A\exp(-E/k_BT),
\end{equation}
where 
\begin{equation}
    A = 1/\sum_E g_kn(E)\exp(-E/k_BT)
\end{equation}
The temperature- and coverage-dependent internal energy can then be calculated as
\begin{equation}
    E_{\theta} = \frac{1}{N}\sum_Eg_kn(E)p^EE,
\end{equation}

and the entropy
\begin{equation}
    S_{\theta} = \frac{-k_B}{N}\sum_Eg_kn(E)p^E\ln(p^E).
\end{equation}
The division by $N$ is to calculate the entropy per surface atom. The free energy $G_{\theta}$ is then calculated by
\begin{equation}
G_{\theta} = E_{\theta} - TS_{\theta}
\end{equation}

Fig. \ref{fig:disorder}b shows the calculated probability distributions for $\theta = 0.5$\,ML at 300 K. The obtained $E_{\theta}$ and $G_{\theta}$ values at 300 K for the disordered phase are plotted in Fig. \ref{fig:disorder}a (red lines). In this way, we construct a thermodynamic model of the disordered surface phase taking into consideration the energy distribution of different solute configurations.

\subsection{Sublattice Model}
We also parameterize and analyze a CALPHAD sublattice model~\cite{Sundman1981, Sundman2018}. Based on the identified surface phases, the model includes three end members: the perfect A surface, the ordered 1/3 A-X surface, and the full X-covered surface. To describe the 1/3 ordered phase, we consider a two-sublattice model (A,X)$_{\sfrac{2}{3}}$(A,X)$_{\sfrac{1}{3}}$. The first sublattice contains the sites occupied by A in the 1/3 ordered phase, and the second sublattice contains the sites occupied by X in the 1/3 ordered phase (denoted here as sublattices 1 and 2). We denote the concentration of A on each sublattice as $c_{\mathrm{A}}^1$ and $c_{\mathrm{A}}^2$, and the concentration of X as $c_{\mathrm{X}}^1$ and $c_{\mathrm{X}}^2$ (here $c_{\mathrm{A}}^1+c_{\mathrm{X}}^1=1$ and $c_{\mathrm{A}}^2+c_{\mathrm{X}}^2=1$). For each sublattice $i$ ($i$ = 1 or 2), a regular solution model is used to describe the free energy of the system:
\begin{eqnarray}
\begin{split}
    G^i = & c_{\mathrm{A}}^iE_{\mathrm{A}}^i + c_{\mathrm{X}}^iE_{\mathrm{X}}^i + w_i c_{\mathrm{A}}^ic_{\mathrm{X}}^i 
\\ & + k_BT[c_{\mathrm{A}}^i\ln c_{\mathrm{A}}^i +c_{\mathrm{X}}^i\ln c_{\mathrm{X}}^i].
\end{split}
\end{eqnarray}

Here $E_{\mathrm{A}}^i$ and $E_{\mathrm{X}}^i$ represent the energy of one A or X atom on the sublattice $i$, respectively. Considering the end members, the values are

\begin{equation}
E_{\mathrm{A}}^1 = E_{\mathrm{A}}^2 = E_{\mathrm{A}},
\end{equation}

\begin{equation}
E_{\mathrm{X}}^1 = (3E_{\mathrm{X}}-3E_{1/3}+2E_{\mathrm{A}})/2,
\end{equation}

\begin{equation}
E_{\mathrm{X}}^2 = 3E_{1/3} - 2E_{\mathrm{A}}.
\end{equation}

Here $E_{\mathrm{A}}$ and $E_{\mathrm{X}}$ represent the energy of the perfect A and X surfaces, normalized per surface atom. Correspondingly, $E_{1/3}$ is the energy of the 1/3 ordered phase. The values of $E_{\mathrm{A}}$, $E_{\mathrm{X}}$, and $E_{1/3}$ depend on the given chemical potentials according to Eq. \ref{eq:surf_energy_general_expression}. The total Gibbs energy of the system is the sum of the Gibbs energies of the two sublattices:
\begin{equation}
    G = \frac{2}{3}G^1 + \frac{1}{3} G^2.
\end{equation}
In this way, when $c_{\mathrm{A}}^1 = c_{\mathrm{A}}^2 = 1$, the system retrieves the energy of the perfect A surface. When $c_{\mathrm{A}}^1 = c_{\mathrm{A}}^2 = 0$, the system energy is that of the full X-covered surface. When $c_{\mathrm{A}}^1 = 1$, $c_{\mathrm{A}}^2 = 0$, the system is in the 1/3 ordered phase and $G = E_{1/3}$. 

Here $G$ is a function of the occupation on the two sublattices, ${c_{\mathrm{A}}^i, c_{\mathrm{X}}^i}$. The surface coverage is  $\theta = \frac{2}{3}c_{\mathrm{X}}^1 + \frac{1}{3}c_{\mathrm{X}}^2$. Therefore a given $\theta$ corresponds to multiple combinations of ${c_{\mathrm{A}}^i, c_{\mathrm{X}}^i}$. We take the $G$ value at given $\theta$ to be the one minimized with respect to ${c_{\mathrm{A}}^i, c_{\mathrm{X}}^i}$:
\begin{equation}
    G_{\theta} = \text{min}\{{G({c_{\mathrm{A}}^i, c_{\mathrm{X}}^i})}\}_{\frac{2}{3}c_{\mathrm{X}}^1 + \frac{1}{3}c_{\mathrm{X}}^2 = \theta}.
\end{equation}
$\theta_{\mathrm{eq}}$ is then the $\theta$ value that minimizes $G_{\theta}$ at the given chemical potentials.

\begin{figure}[t] 
\centering
\includegraphics[width=0.4\textwidth]{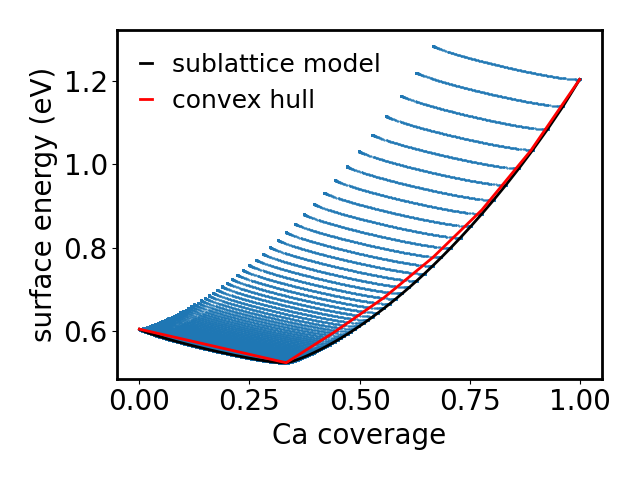}
\caption{\label{fig:regular-solution} The sublattice model for the Mg-Ca system (black line). The blue dots are the energies of surface structures with different discrete combinations of \{$c_{\mathrm{A}}^i, c_{\mathrm{X}}^i$\}. The red line is the original convex hull from Fig. \ref{fig:disorder}a. }
\end{figure}

The non-ideal mixing parameter $w_i$ is critical in determining the ordering of solute atoms. We systematically optimize the $w_i$ values until the resulting segregation isotherm best fits those obtained from SGCMC calculation. For the Mg-Ca system, we use $w_1 = -0.30$ and $w_2 = 0.02$ eV, respectively (see details in SM~\cite{SM}). The corresponding energy profile is plotted in Fig. \ref{fig:regular-solution}.

\section{Results and Discussion}

In this section, we first discuss the equilibrium surface structures and surface segregation isotherms obtained with SGCMC. We then apply our proposed SPEA model and test its performance as a numerically efficient surrogate model for the expensive SGCMC calculations which require energy evaluations of millions of structure for any point in the phase diagram. Finally, we compare the SPEA model with the sublattice model, an alternative analytical model widely used in CALPHAD modelingto build phase diagrams.

\subsection{Semi-grand Canonical Monte Carlo}

\begin{figure}[h] 
\centering
\includegraphics[width=0.5\textwidth]{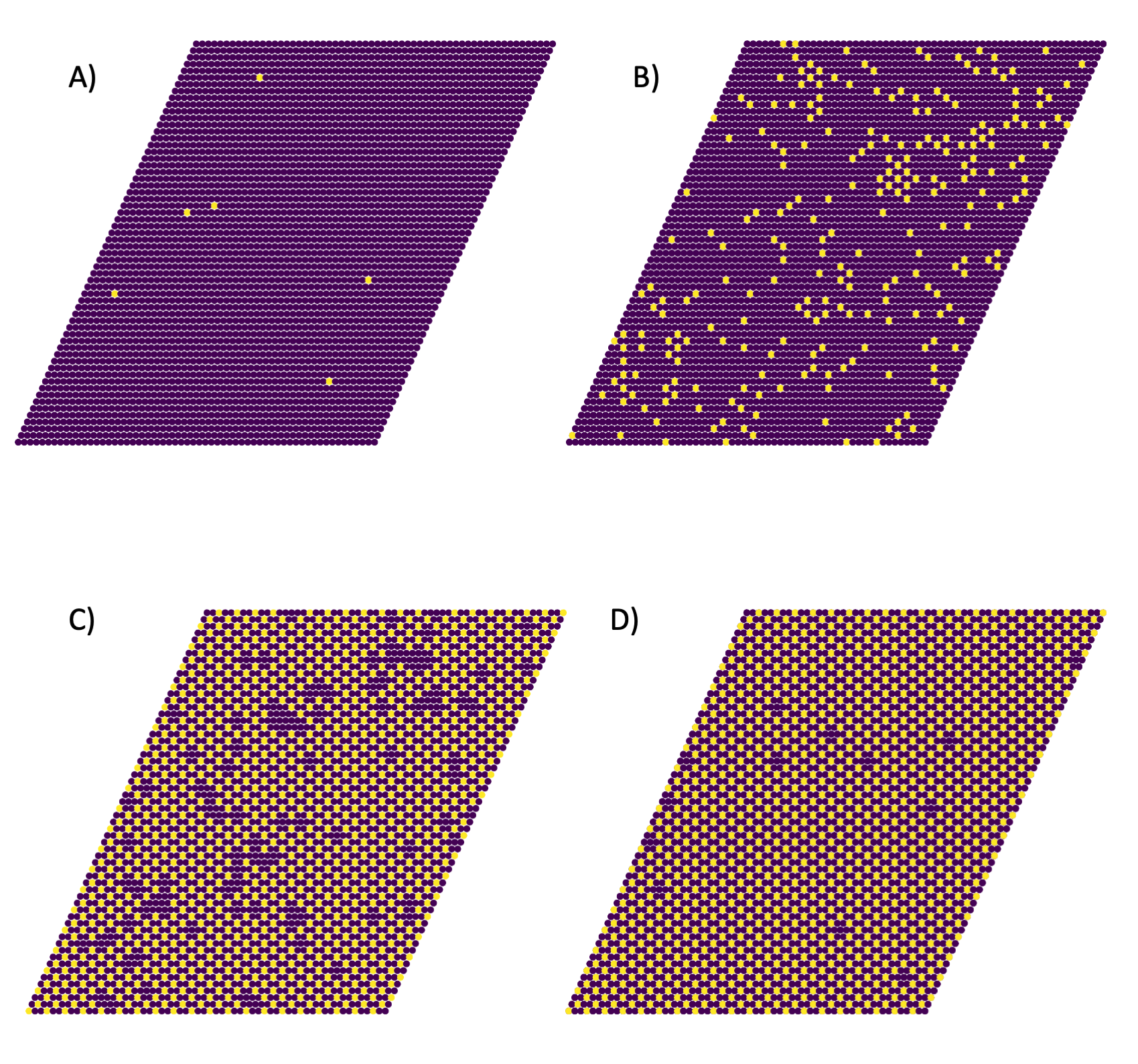}
\caption{\label{fig:MC-Mg-Ca} Surface structures obtained by semi-grand canonical Monte Carlo (SGCMC) simulations of the Mg-Ca surface at different Ca chemical potentials $\mu_{\rm Ca}$. The purple and yellow spheres represent Mg and Ca atoms, respectively.   }
\end{figure}

\begin{figure}[h] 
\centering
\includegraphics[width=0.5\textwidth]{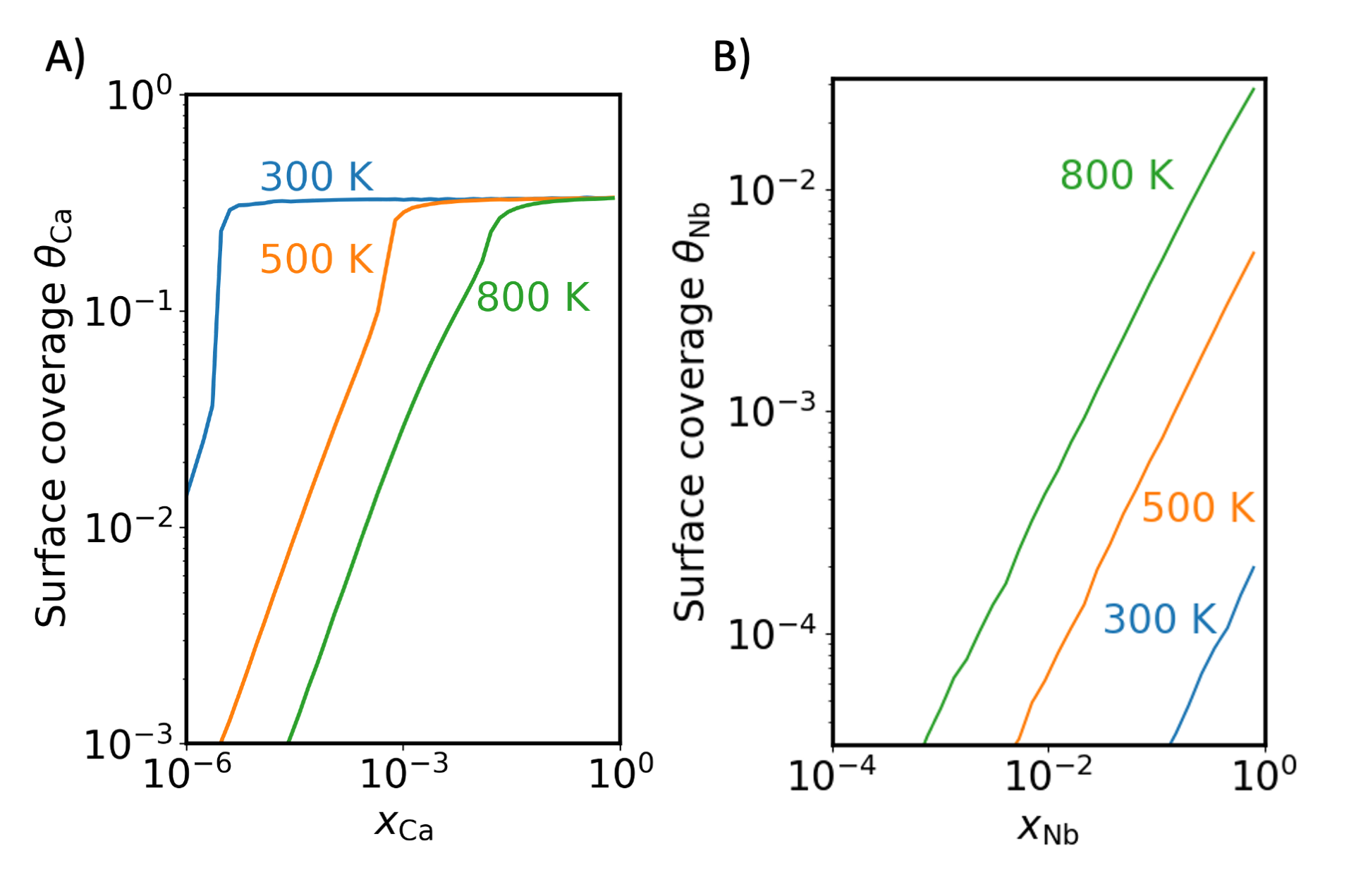}
\caption{\label{fig:MC-isotherm} Surface segregation isotherms for (a) the Mg-Ca system and (b) the Ni-Nb system obtained with SGCMC at different temperatures. The chemical potential $\mu_{\mathrm{X}}$ at $x_{\mathrm{X}} = 1$ is equal to $E_{\mathrm{X\text{-}in\text{-}A\text{-}bulk}}$ according to Eq. \ref{eq:chemical_potential}.}
\end{figure}

We first show the surface structures and segregation isotherms obtained with SGCMC for the Mg-Ca system. Fig. \ref{fig:MC-Mg-Ca}  shows how the surface configurations change with increasing $\mu_{\mathrm{Ca}}$ in the Mg-Ca system. Under Ca-poor conditions, i.e. low $\mu_{\mathrm{Ca}}$, the segregated Ca atoms are randomly distributed on the surface (Fig. \ref{fig:MC-Mg-Ca}a). With increasing $\mu_{\mathrm{Ca}}$, the concentration of segregated Ca also increases, and we observe the formation of islands with the ordered 1/3 structure (Fig. \ref{fig:MC-Mg-Ca}b). Further increase of $\mu_{\mathrm{Ca}}$ gradually leads to the filling of the 1/3 structure, with only a few defects on the Ca sublattice (Fig. \ref{fig:MC-Mg-Ca}c and d). This is also shown by the surface isotherm in Fig. \ref{fig:MC-isotherm}a, where we observe this transition from the disordered surface phase to the ordered surface phase at different temperatures. The disordered phase shows a continuous increase in the surface coverage with increasing $\mu_{\mathrm{Ca}}$ and the ordered phase a plateau.  

\begin{figure}[h] 
\centering
\includegraphics[width=0.4\textwidth]{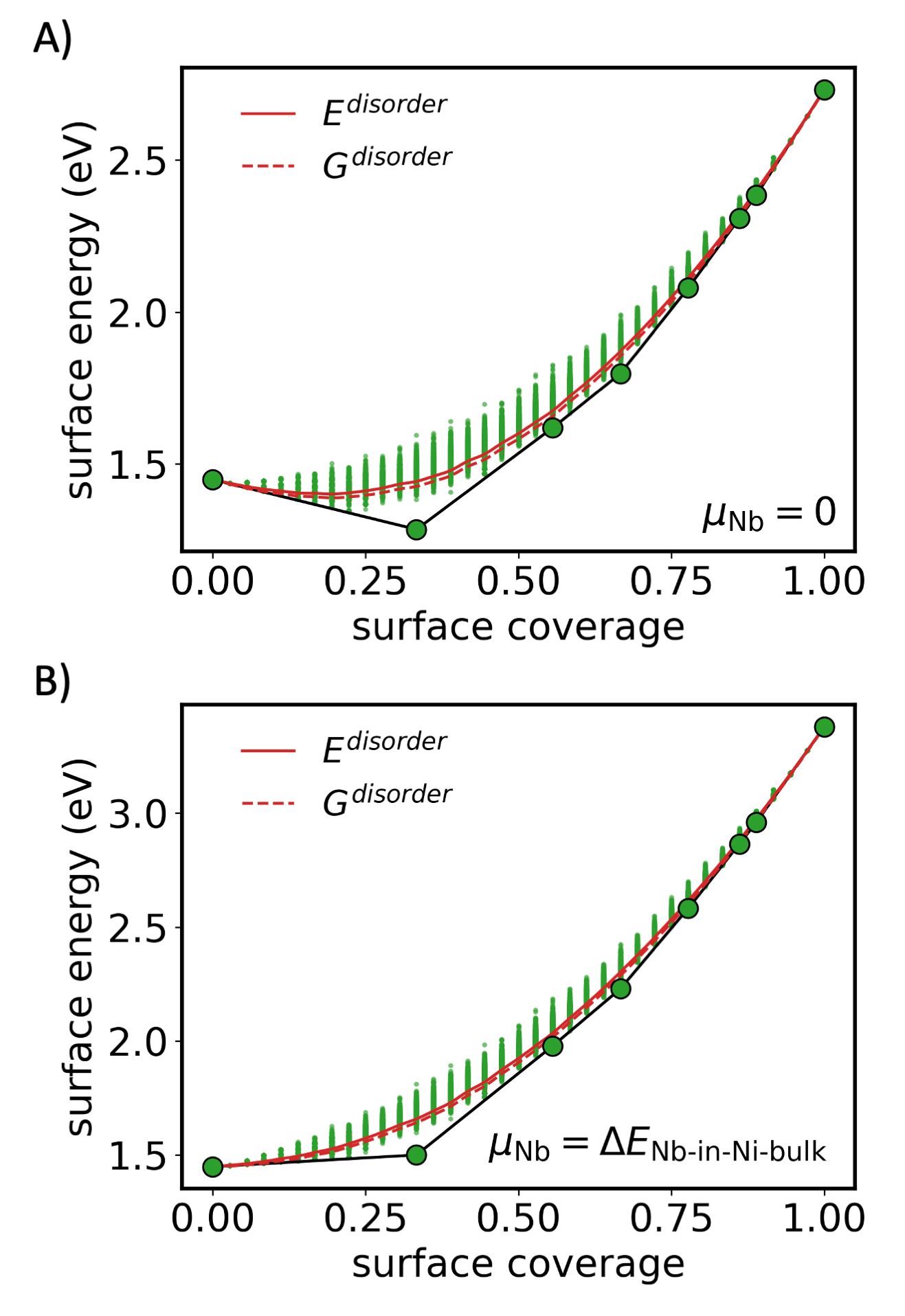}
\caption{\label{fig:ni-nb} Energy distribution of Ni-Nb surface structures with different solute configurations  as a function of the surface solute coverage $\theta$ predicted by CE (a) at the chemical potential of bulk Nb metal and (b) at the chemical potential of one Nb atom in bulk Ni. The surface energy is shown for (1$\times$1) surface unit cell. The chemical potential of Ni is assumed to be that of Ni bulk. The $E^{disorder}$ and $G^{disorder}$ lines are the energy and free energy of the disordered phase at 300\,K calculated from the thermodynamic model described in Sec. IIC.  }
\end{figure}

The surface segregation isotherm for the Ni-Nb system is distinct from that of Mg-Ca despite the similarity of the convex hulls of the two systems. Instead of showing a transition from the disordered phase to the 1/3 ordered phase, the disordered phase prevails even at high bulk Nb concentrations (Fig. \ref{fig:MC-isotherm}b). Fig. \ref{fig:ni-nb} shows the convex hull of the Ni-Nb surface as predicted by CE. Looking at the convex hull plot referenced to the chemical potentials of the bulk metals, both the Mg-Ca (Fig. \ref{fig:disorder}a) and Ni-Nb (Fig. \ref{fig:ni-nb}a) systems show a stable 1/3 ordered structure. The energy of the ordered phase is separated from the disordered structures by an energy gap, which makes the ordered phase thermodynamically stable against disordering until the thermal energy becomes so large that it can overcome the energy gap. For the Ni-Nb system, the ordered 1/3 phase is never reached. This is because of the formation of a competing bulk intermetallic phase. The chemical potential at which the disorder-order transition happens is higher than the thermodynamic upper limit of $\mu_{\mathrm{Nb\text{-}in\text{-}Ni\text{-}bulk}}$ and therefore does not show up in the $\theta$-$x_{\mathrm{Nb}}$ plot. For more details on the thermodynamic limit of $\mu_{\mathrm{Nb}}$ in bulk Ni, see SM~\cite{SM}.

In the remainder of this section, we will use the surface segregation isotherms obtained with SGCMC as a strict benchmark for evaluating the performance of the analytical thermodynamic models. It should be noted that in the low coverage limit, the SGCMC accuracy is limited by the cell size and sampling statistics, so we only focus on the region where the coverage is greater than 10$^{-3}$. As we are interested in constructing phase diagrams, which are characterized by phase boundaries, and the phase transitions occur typically only at concentrations well above this value, the available concentration accuracy is more than sufficient. Moreover, in the low coverage limit, solute-solute interaction  is negligible and can be safely ignored. In this limit, the solute binding energy is given by a single value, i.e. that of an isolated solute in the host matrix. This can be described by the McLean model~\cite{McLean1957}. The McLean model leads to a straight line in double logarithmic coverage versus bulk concentration plot, which is what we observe in Fig. \ref{fig:MC-isotherm} in the low concentration region. The linearity is direct evidence of the validity of the McLean model, i.e. of negligible solute-solute interaction.

\subsection{The SPEA method}

The SGCMC results in Fig. \ref{fig:MC-Mg-Ca} show that the phase transition is not abrupt but extends over a limited interval of chemical potential. In this region, phase coexistence, i.e. the simultaneous presence of two surface phases, is observed. This observation motivates us to construct a model, in which the distribution of the different surface phases is given by Boltzmann weighting. A similar method has previously been used to compute the thermodynamic properties of multi-component or strongly disordered bulk materials~\cite{Yang2016,Liu2024}. Mathematically, this can be expressed by the partition function $Z$ of the system 
\begin{equation}
Z = \sum_{\theta} \exp(-NG_{\theta}/k_BT).
\end{equation}
Here, the subscript $\theta$ represents the different solute concentrations on the surface. The coefficient $N$ is a scaling factor. It represents the minimum number of surface atoms a surface phase has to have to bethermodynamically stable (see Discussion). The area fraction, i.e., the area a surface phase has with regard to the total surface area is then
\begin{equation}
    p^{\theta} = \exp(-NG_{\theta}/k_bT)/Z.
\end{equation}
The equilibrium solute concentration on the surface is the sum of all possible phases according to their probabilities $p^{\theta}$:
\begin{equation}
    \theta_{\mathrm{eq}} = \sum p^{\theta} \theta.
\end{equation}

In Sec. IID, we have given a detailed description of the sublattice model. Here we compare the surface segregation isotherms obtained with the SPEA model and the sublattice model, and benchmark them against the SGCMC results. For the critical island size $N$ in the SPEA model, we find $N = 10$ to provide a good description of all segregation isotherms. For the sublattice model, we find $w_1$ = $-$0.3, $w_2$ = 0.02 eV for the Mg-Ca system and  $w_2$ = 0.01 eV for the Ni-Nb system to provide an optimal fit (see further details in Discussion). 

\begin{figure}[h] 
\centering
\includegraphics[width=0.5\textwidth]{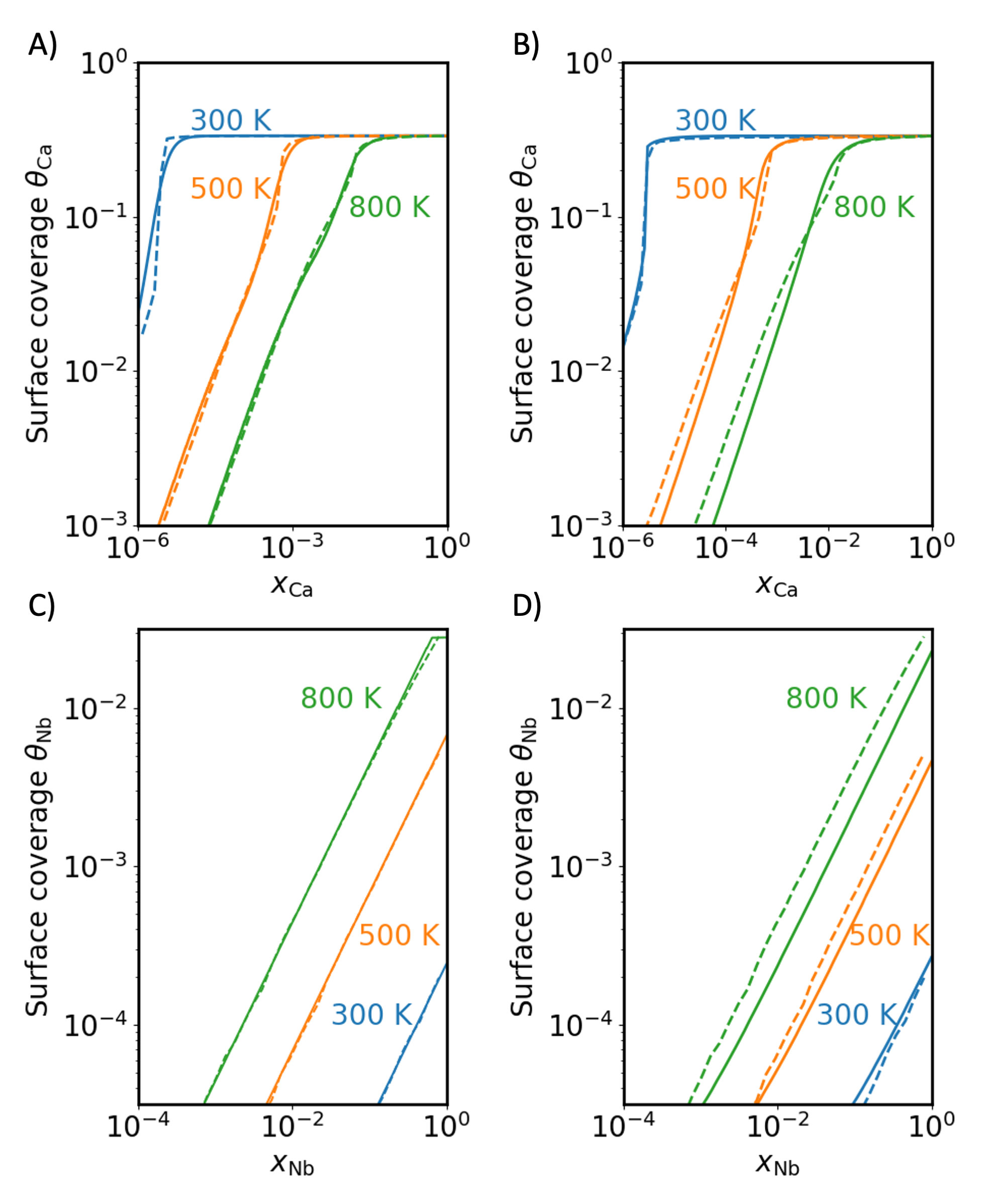}
\caption{\label{fig:analytical_isotherm} Surface segregation isotherms for the Mg-Ca system calculated with (a) the SPEA model and (b) the sublattice model and for the Ni-Nb system calculated with (c) the SPEA model and (d) the sublattice model. The dashed lines represent the SGCMC result at the corresponding temperatures as shown in Fig. \ref{fig:MC-isotherm}. }
\end{figure}

Both models are able to produce segregation isotherms that agree well with the SGCMC results (Fig. \ref{fig:analytical_isotherm}). In the case of Mg-Ca (Fig. \ref{fig:analytical_isotherm} a and b), both models predict the critical concentration $x_{\mathrm{Ca}}$ at which the order-disorder transition occurs with good accuracy across the whole temperature range. In the transition zone, the SPEA model matches the SGCMC curve slightly better, especially at higher temperatures. This behavior is expected because the SPEA model takes into account a mixture of the disordered phase and the ordered phase, which more accurately reflects the surface morphology observed in the SGCMC simulation (Fig. \ref{fig:MC-Mg-Ca}). In the low-concentration disordered range, all models show a linear surface coverage versus bulk concentration ($\theta_{\mathrm{Ca}}$-$x_{\mathrm{Ca}}$) behavior in the log-log plot, which is expected for ideal mixing. The SPEA model shows very good agreement with the SGCMC result in this region, while the sublattice model shows a slight quantitative deviation. The same deviation is observed for the Ni-Nb system (Fig. \ref{fig:analytical_isotherm} d). This deviation is likely due to the deviation from the asymptotic behavior at $\theta_{\mathrm{Ca}}\rightarrow0$ of the $G-\theta_{\mathrm{Ca}}$ curve. A better quantitative agreement can be achieved by fine-tuning the parameters $N$/$w_i$ used for the model at each temperature and chemical potential. This is further discussed in Sec. IIIC.           

\subsection{Discussion}
By comparing the surface segregation isotherms predicted by SGCMC simulation and the two thermodynamic models, we show that both models give sufficiently accurate results. Here we further discuss the advantages of the SPEA model over the sublattice model. 

\begin{figure}[h] 
\centering
\includegraphics[width=0.5\textwidth]{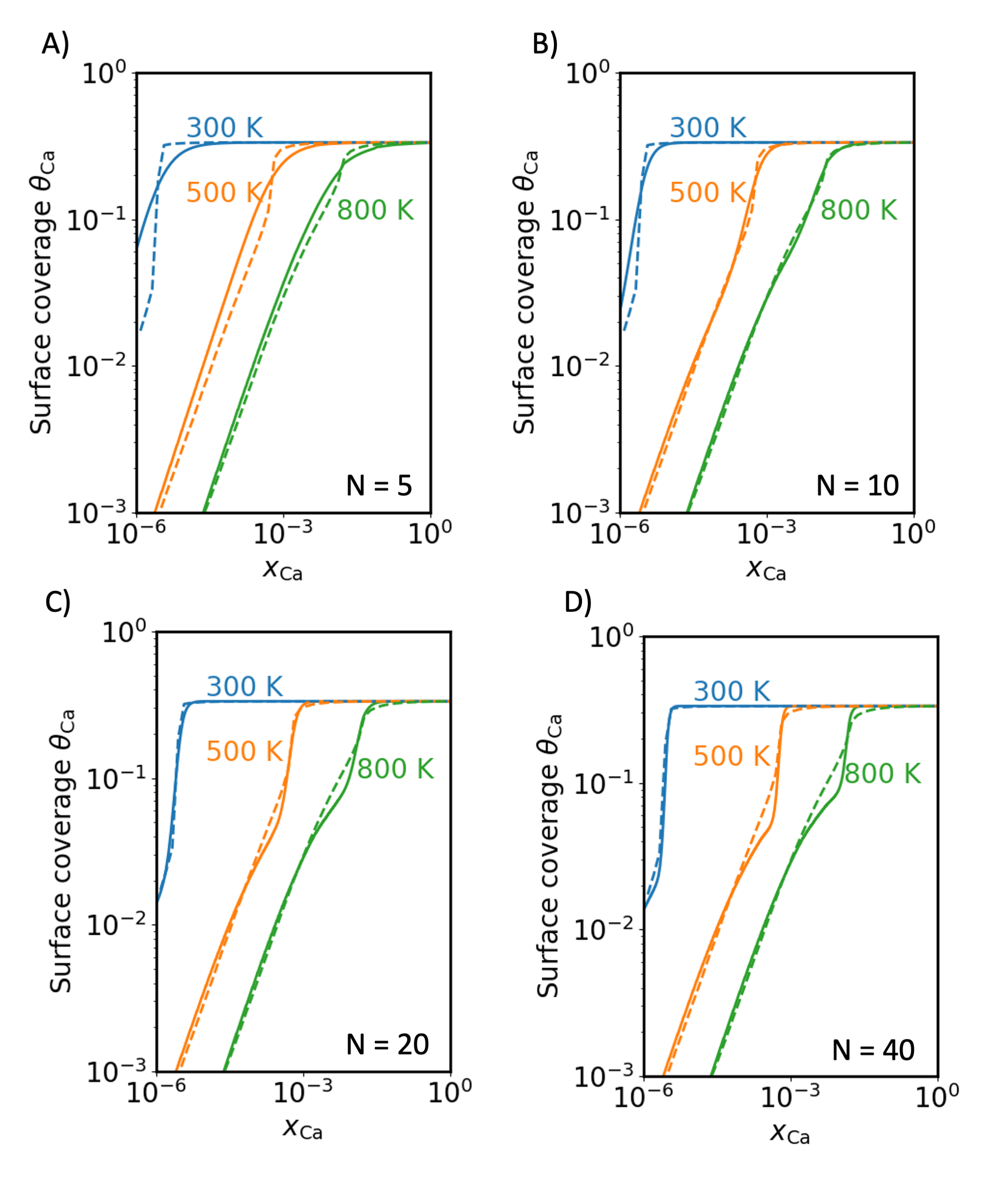}
\caption{\label{fig:isotherm_N} Surface segregation isotherms of the Mg-Ca system obtained with the SPEA model at different values of $N$. The dashed lines represent the SGCMC result at the corresponding temperatures as shown in Fig. \ref{fig:MC-isotherm}. }
\end{figure}

\begin{figure}[h] 
\centering
\includegraphics[width=0.5\textwidth]{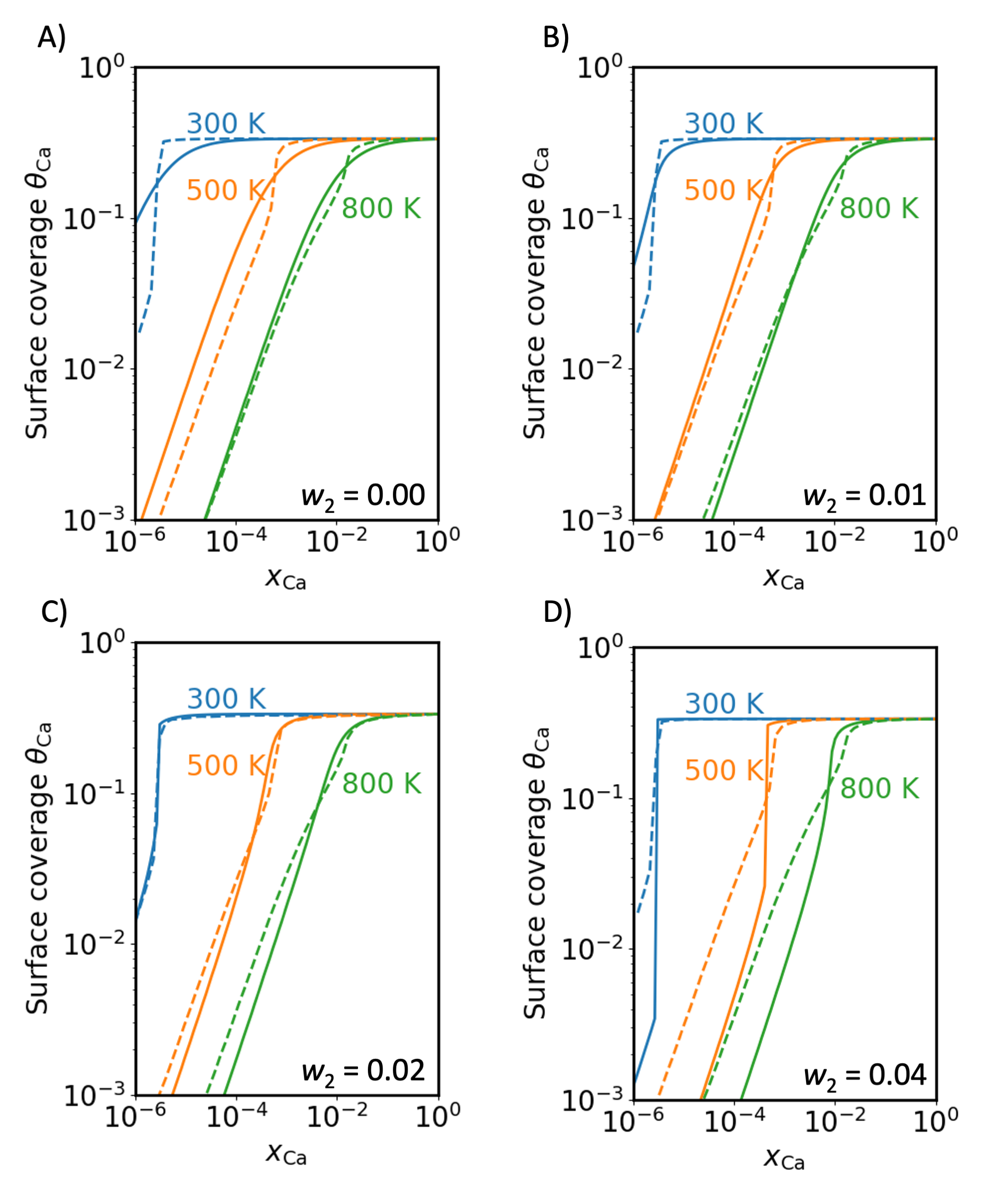}
\caption{\label{fig:isotherm_w} Surface segregation isotherms of the Mg-Ca system obtained with the sublattice model at different values of $w_2$. The dashed lines represent the SGCMC result at the corresponding temperatures as shown in Fig. \ref{fig:MC-isotherm}. }
\end{figure}

\begin{figure}[h] 
\centering
\includegraphics[width=0.3\textwidth]{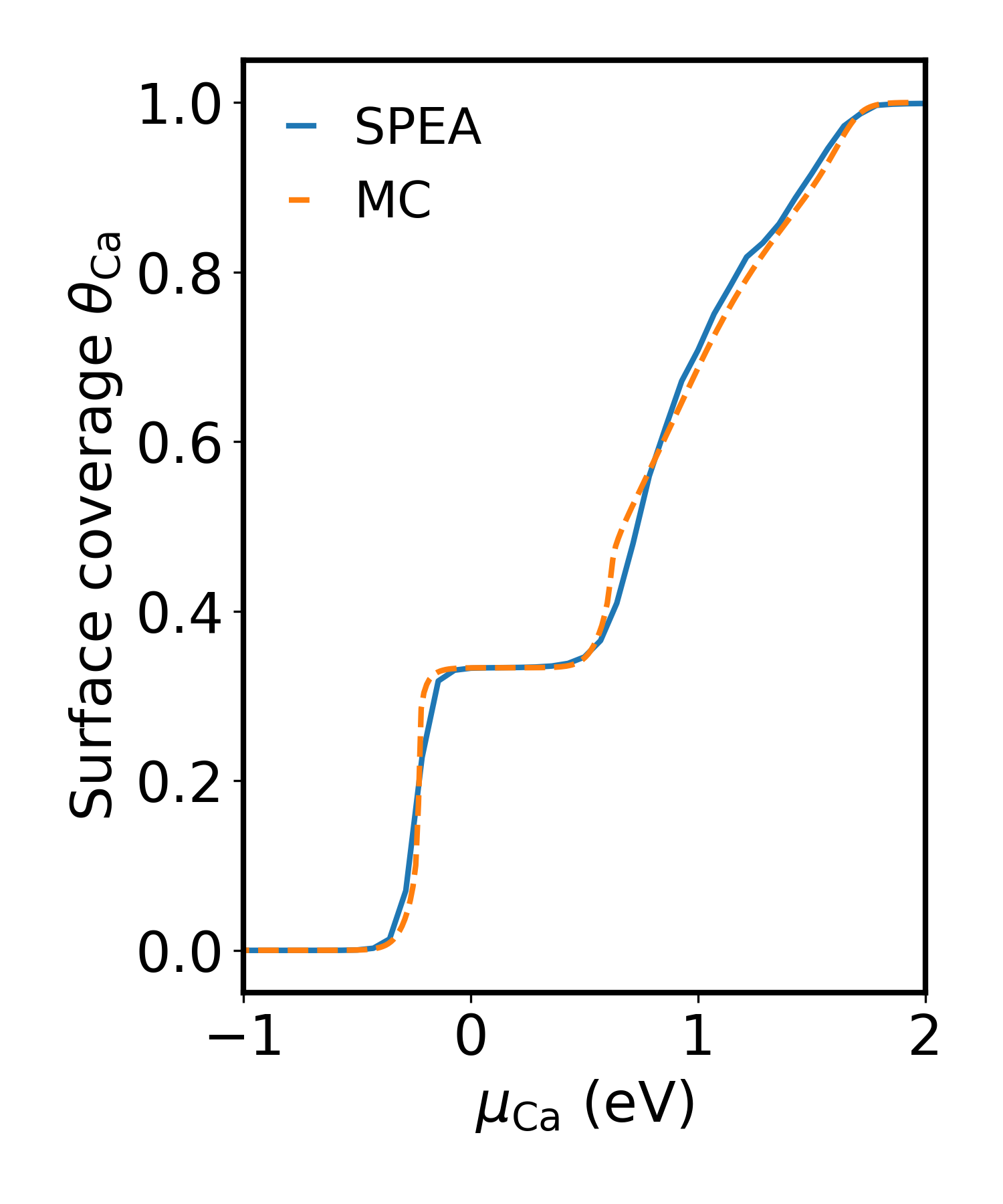}
\caption{\label{fig:mu} Surface segregation isotherm of the Mg-Ca system in the extended $\mu_{\mathrm{Ca}}$ space at 500 K. }
\end{figure}

First, although both models include empirical parameters ($N$ for the SPEA model and $w_i$ for the sublattice model), the SPEA model is much less sensitive to the choice of the parameter. We demonstrate this parameter dependence in Figs. \ref{fig:isotherm_N} and \ref{fig:isotherm_w} for the Mg-Ca system. In Fig. \ref{fig:isotherm_N}, we show the surface segregation isotherms calculated with $N = $ 5, 10, 20, and 40. We observe that the value of $N$ only affect the abruptness of the transition between the disordered and ordered phases. At $N =$ 5, the system undergoes a smooth transition between the two phases and at $N = 40$, the transition is almost vertical. Apart from this change, the isotherm remains largely identical. Notably, the optimal critical island size has a monotonous dependence on temperature. For 300, 500, and 800 K, the optimized $N$ values are 27, 19, 11, respectively. This reflects the thermodynamic reality that the probability of forming islands with larger sizes increases when the temperature is lowered. However, in general, the error changes only slightly with the choice of $N$. See the error versus $N$ plot in SM~\cite{SM}. 

The sublattice model, however, depends quite strongly on the choice of $w_2$. In Fig. \ref{fig:isotherm_w}, we show the surface segregation isotherms calculated with $w_2 = $ 0.00, 0.01, 0.02, and 0.04. Here we observe that although the chemical potential at which the order-disorder transition occurs does not critically depend on $w_2$, the predicted isotherm changes substantially in both the low-coverage disordered region and the transition region. For small $w_2$, the equilibrium coverage in the disordered phase is overestimated and for large $w_2$ significantly underestimated. This difference in parameter dependence leads to an advantage in using the SPEA model in practice. It removes the need to optimize the value of $N$ for different systems. Here we see that $N$ = 10 works reasonably well for both the Mg-Ca and the Ni-Nb systems, and the result is robust to changes of $N$. However, for the sublattice model it is necessary to optimize the $w_i$ values for each individual system. Here $w_2$ = 0.02 works well for the Mg-Ca system but gives a rather significant error for the Ni-Nb system. 

Second, when multiple ordered phases exist, the complexity of the sublattice model increases significantly while the SPEA model does not. Adding end members to the sublattice model also increases the number of sublattices and the subsequent $w_i$. This increase in complexity combined with the sensitivity of the model to $w_i$ values makes it substantially harder to parameterize the model. On the other hand, the SPEA model, by construction, can very easily include multiple ordered phases without increasing the complexity of the model. To demonstrate this, we show the surface segregation isotherm of the Mg-Ca system in the extended $\mu_{\mathrm{Ca}}$ space in Fig. \ref{fig:mu} and compare with MC results. Here the range of $\mu_{\mathrm{Ca}} >$  95 meV corresponds to extreme conditions with the chemical potential above the thermodynamic limit in bulk Mg. While such high chemical potential is out of reach in realistic scenarios, it allows us to cover a large set of possible surface phases. As shown in Fig. \ref{fig:mu}, after a plateau of the 1/3 ordered phase, the surface coverage continues to rise until it reaches full saturation. This transition zone corresponds to a mixture of all the ordered phases on the right side of the convex hull with $\theta>$ 1/3 (Fig. \ref{fig:disorder}) and also the disordered phase. By comparing the SPEA model with our reference MC data, we find a highly consistent agreement over the whole $\mu_{\mathrm{Ca}}$ range.


Given these considerations, we propose the SPEA model as an accurate surrogate model for the computationally much more expansive MC simulations. It provides a detailed understanding of the evolution of surface equilibrium structures and serves as an efficient and highly accurate approach for constructing surface phase diagrams. It can qualitatively and quantitatively reproduce the result of MC simulations with satisfactory accuracy, and is robust to changes of the empirical parameter $N$. The SPEA model works particularly well in combination with high-throughput \textit{ab initio} computing, where one can scan the configuration space and calculate the energy of each surface structure. Compared to the sublattice model, it eliminates the need to identify the end members and parameterize the model for each sublattice. Instead, the SPEA model predicts the fraction of each possible surface phase and provides information on the phase mixing. The method can be easily extended to the construction of other types of phase diagrams and potentially serve as a more effective approach for incorporating \textit{ab initio} data into CALPHAD modeling.

\section{Conclusion}

In conclusion, we have proposed an analytical model for efficiently computating defect phase diagrams, the SPEA model. It assumes a Boltzmann distribution of the phase fractions and is thus capable of reproducing the defect phase transitions where phase separation occurs over a finite interval of chemical potentials. We demonstrate the performance of the proposed analytical model on reproducing the surface segregation of Mg-Ca and Ni-Nb alloys. The model accurately captures the mixture of the ordered 1/3 coverage phase and the disordered phase, as observed in the SGCMC simulation. The SPEA model has advantages over an alternative analytical model, the CALPHAD sublattice model, in requiring less empirical parametrization and providing more insights into the phase fractions. It thus serves as a computationally efficient method for replacing MC thermodynamic averages and is extendable to the construction of generic defect phase diagrams independent of the defect type or material system.


\begin{acknowledgments}
We acknowledge funding by the Deutsche Forschungsgemeinschaft (DFG, German Research Foundation) through SFB1394, project no. 409476157 and SFB1625, project no.506711657.
\end{acknowledgments}


\bibliography{reference}

\end{document}


\preprint{APS/123-QED}

\title{Supplemental Material for\\ Order-disorder transitions on alloy surfaces: Monte Carlo thermodynamic averages versus analytical models}

\author{Jing Yang}
\email{j.yang@mpie.de}
\affiliation{Computational Materials Design Department, Max Planck Institute for Sustainable Materials, Max-Planck-Str. 1, D-40237 Düsseldorf, Germany}

\author{Ahmed Abdelkawy}
\affiliation{Computational Materials Design Department, Max Planck Institute for Sustainable Materials, Max-Planck-Str. 1, D-40237 Düsseldorf, Germany}

\author{Mira Todorova}
\email{m.todorova@mpie.de}
\affiliation{Computational Materials Design Department, Max Planck Institute for Sustainable Materials, Max-Planck-Str. 1, D-40237 Düsseldorf, Germany}
 
\author{Jörg Neugebauer}%
\affiliation{Computational Materials Design Department, Max Planck Institute for Sustainable Materials, Max-Planck-Str. 1, D-40237 Düsseldorf, Germany}

\maketitle
\beginsupplement

\section{The cell size convergence of the Monte Carlo calculation}

\begin{figure}[h] 
\centering
\includegraphics[width=0.5\textwidth]{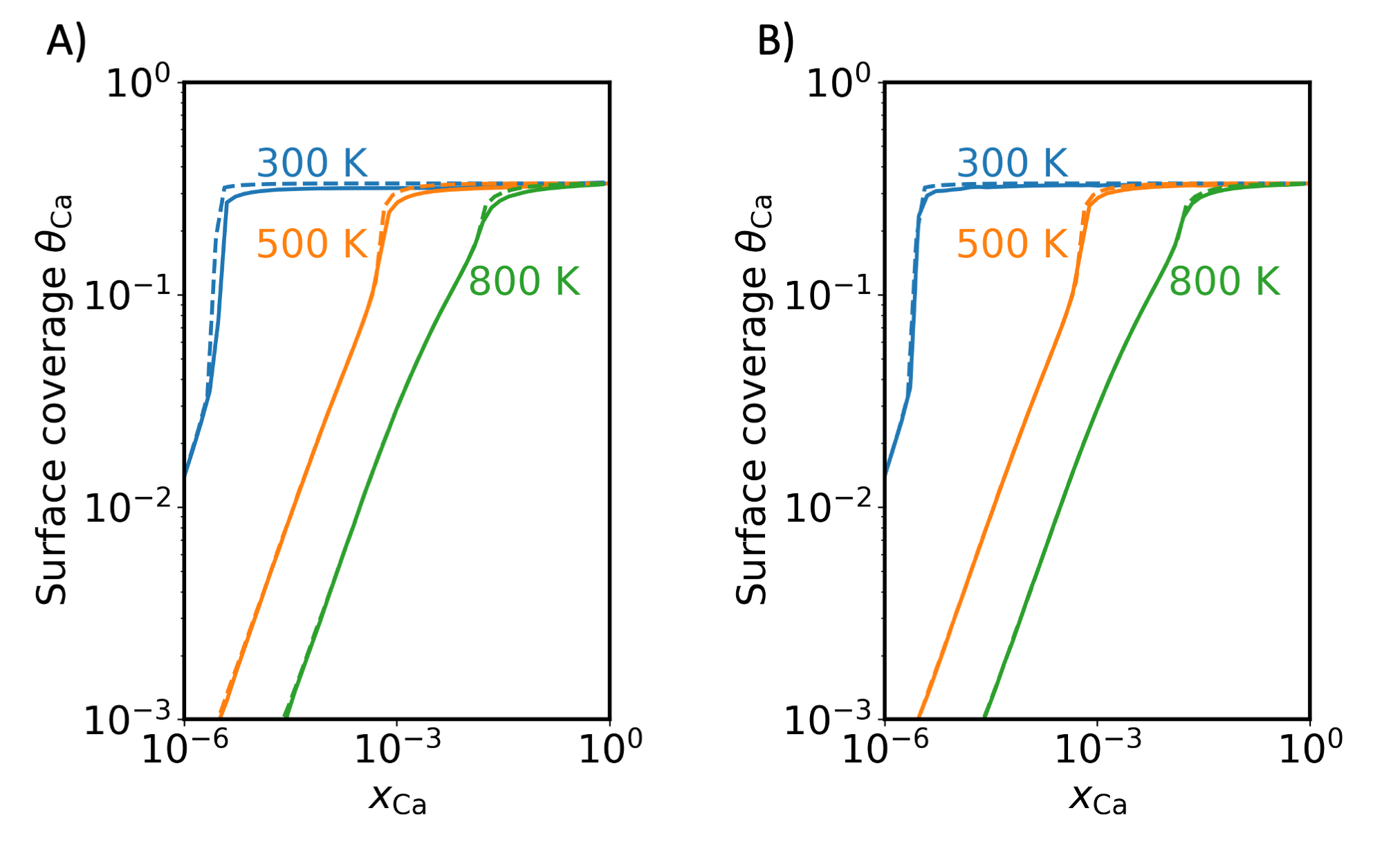}
\caption{\label{fig:cell_size} SGCMC surface segregation isotherm of the Mg-Ca system calculated with (a) (40$\times$40) surface cell and (b) (80 $\times$ 80) surface cell. The dashed lines represent the result of using a (60$\times$60) surface cell as reported in the main article. }
\end{figure}

We observe very limited effect of the cell size used in the SGCMC calculation. In the main article, we report the results of using a (60$\times$60) surface cell (3600 surface atoms). Fig. \ref{fig:cell_size} shows the results of using (40$\times$40) and (80 $\times$ 80) surface cells. Apart from a slight change in the transition zone, the different cell sizes give almost identical results and do not affect the comparison between the different models in the main article.  

\section{Parameter selection for the sublattice model}

\begin{figure}[h] 
\centering
\includegraphics[width=0.4\textwidth]{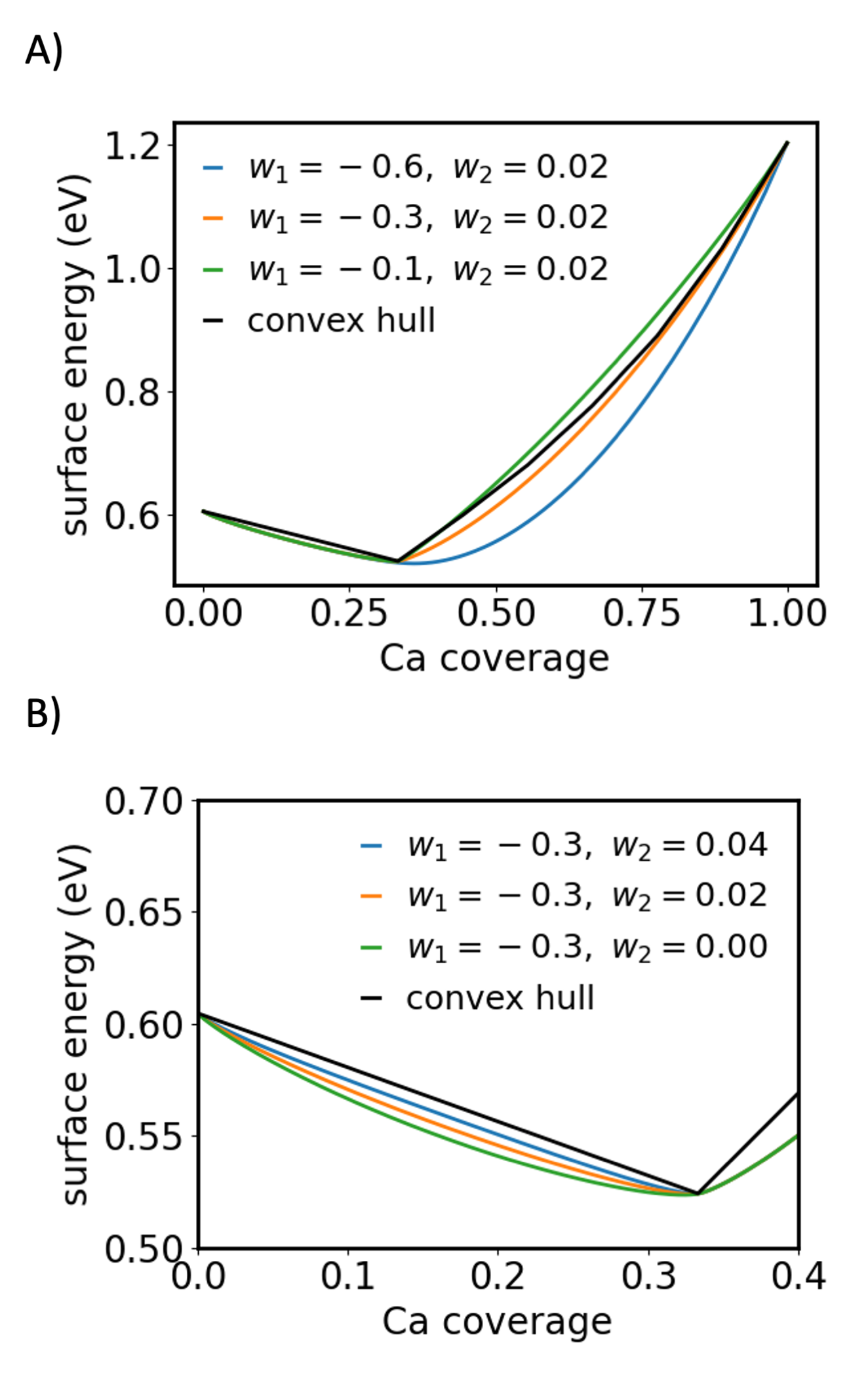}
\caption{\label{fig:w_effect} The $G-\theta_{\mathrm{Ca}}$ function calculated with the sublattice model at 800\,K with (a) different $w_1$ values and (b) different $w_2$ values. The energies are calculated at the chemical potential of bulk Mg and Ca metal. }
\end{figure}

The parameters $w_1$ and $w_2$ significantly influence the behavior of the sublattice model. In Fig. \ref{fig:w_effect}, we show the $G$ values calculated with the sublattice model as a function of the Ca coverage $\theta_{\mathrm{Ca}}$ at different $w_1$ and $w_2$ values. By the given definition, the value of $w_1$ affects the $G$ vs. $\theta_{\mathrm{Ca}}$ curve in the range where $\theta_{\mathrm{Ca}} > 1/3$\,ML, and $w_2$ affects the range where $\theta_{\mathrm{Ca}} < 1/3$\,ML. The magnitude of $w_i$ determines the curvature of the curve. Since here the surface coverage never exceeds 1/3\,ML in the relevant chemical potential range, the value of $w_1$ does not have an effect on the presented results in the main article and is not optimized. The value of $w_2$ affects both the equilibrium coverage in the disordered region and the transition region as discussed in the main article.

\section{Surface ordering of the Ni-Nb system }

\begin{figure}[h] 
\centering
\includegraphics[width=0.35\textwidth]{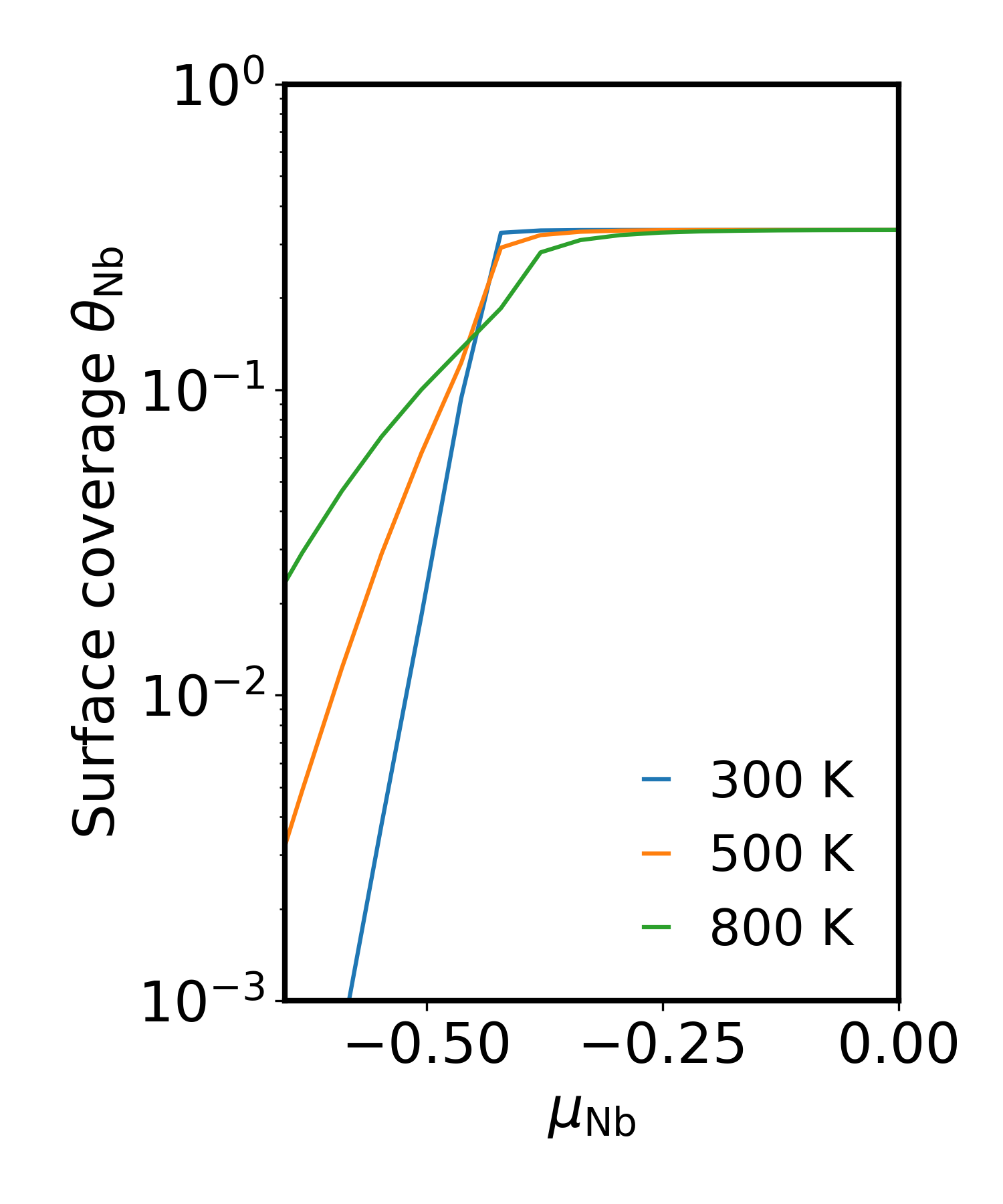}
\caption{\label{fig:isotherm_mu} The surface segregation isotherm of the Ni-Nb system as a function of the chemical potential of Nb. }
\end{figure}

In the Ni-Nb system, although the 1/3\,ML ordered phase is an energetically stable phase, it does not form in the thermodynamic range where the Ni-Nb solid solution is stable. We show the transition from the disordered phase to the ordered phase in the $\mu_{\mathrm{Nb}}$ in Fig. \ref{fig:isotherm_mu}. Here we can see that the transition occurs at about $\mu_{\mathrm{Nb}}$ = $-$0.4 eV. This chemical potential is higher than the value of $\Delta E_{\rm Nb\text{-}in\text{-} Ni\text{-}bulk}$, which is the upper limit of $\mu_{\mathrm{Nb}}$ in the Ni-Nb solid solution. At $\mu_{\mathrm{Nb}}$ = $-$0.4 eV, the precipitation of Ni-Nb intermetallics is expected. 

In the main article, we limit our discussion to the thermodynamic range where the solid solution is thermodyanmically stable. Therefore we do not observe the order-disorder transition in the Ni-Nb system. However, the method described in the main article works for an extended range of $\mu_{\mathrm{Nb}}$, and can potentially describe the surface phase formation in the metastable region.  

\begin{figure}[h] 
\centering
\includegraphics[width=0.5\textwidth]{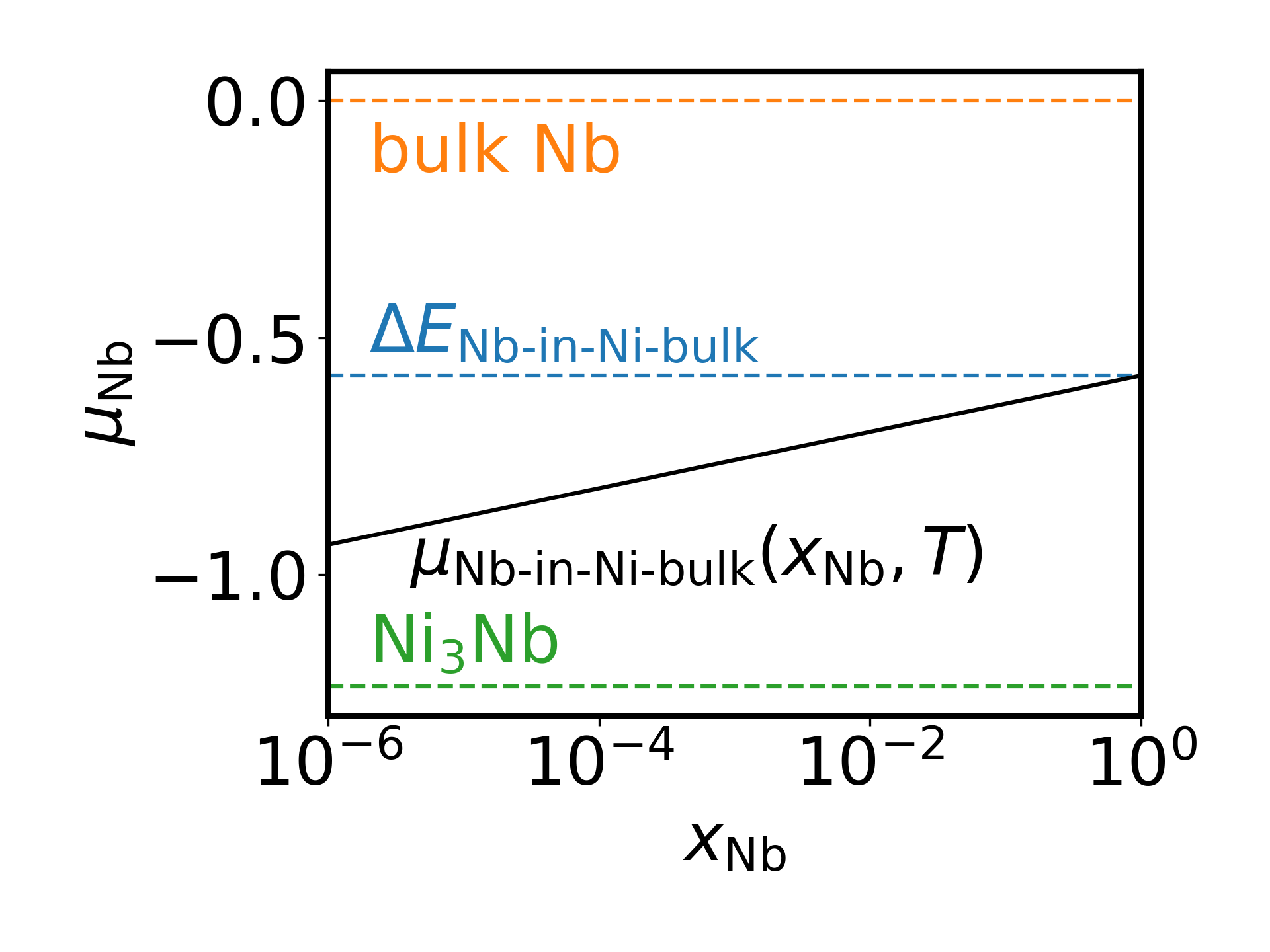}
\caption{\label{fig:Ni_Nb_chemical_potential} The chemical potential of Nb in Ni as a function of its concentration $x_{\mathrm{Nb}}$ at 300 K plotted against the two reference phases, bulk Nb and the Ni$_3$Nb intermetallic phase.  }
\end{figure}

In Fig. \ref{fig:Ni_Nb_chemical_potential}, we show the chemical potential of Nb in Ni $\mu_{\mathrm{Nb\text{-}in\text{-}Ni\text{-}bulk}} = \Delta E_{\mathrm{Nb\text{-}in\text{-}Ni\text{-}bulk}} + k_BT\log x_{\mathrm{Nb}}$ (black line) and compare it with the two reference phases, bulk Nb and the Ni$_3$Nb intermetallic phase. We can see that the upper limit of $\mu_{\mathrm{Nb\text{-}in\text{-}Ni\text{-}bulk}}$ is -0.65 eV, which is still above the transition chemical potential in Fig. \ref{fig:isotherm_mu}. Therefore, in the $\theta$-$x_{\mathrm{Nb}}$ plot, we do not see any phase transformation.

\section{The critical island size $N$}
\begin{figure}[h]
\centering
\includegraphics[width=0.5\textwidth]{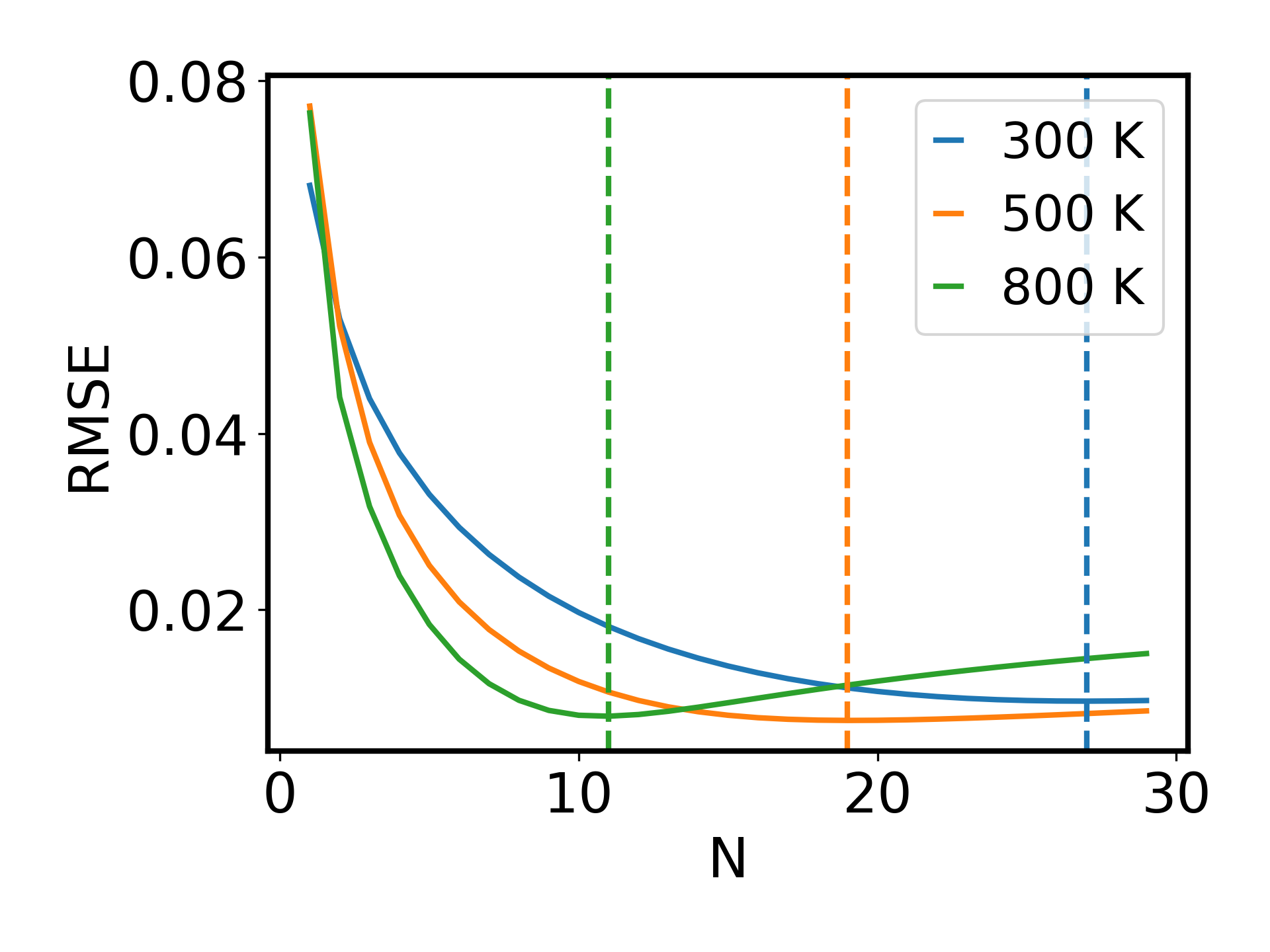}
\caption{\label{fig:N_RMSE} The root mean square error of the segregation isotherm predicted with the SPEA model benchmarked against the MC results, as a function of the critical island size $N$. The vertical lines show the $N$ values corresponding to the error minimum at each temperature. }
\end{figure}
In Fig. \ref{fig:N_RMSE}, we show how the error of the SPEA model changes with the choice of the critical island size $N$. The root mean square error (RMSE) is calculated against the isotherms obtained from the MC calculation. Here it is clear that the optimal $N$ value increases with decreasing temperature ($N$ = 27, 19, 11 for 300 K, 500 K, and 800 K). However, except for the region where $N$ becomes unphysically small, the dependence of the RMSE on $N$ is relatively weak. As we have seen in the main manuscript, using $N$ value between 10 and 30 produces results that are both qualitatively and quantitatively very similar.
